\documentclass[sigconf, nonacm]{acmart}
\usepackage[titletoc,title]{appendix}
\usepackage{gensymb}
\usepackage{fancyhdr}
\usepackage{lipsum}

\newcommand\vldbdoi{XX.XX/XXX.XX}

\newcommand\vldbavailabilityurl{http://vldb.org/pvldb/format_vol14.html}

\begin{document}
\rfoot{Page \thepage}


\title{A Framework for Auditing Data Center Energy Usage and Mitigating Environmental Footprint}

\author{Justin A. Gould}
\affiliation{%
  \institution{Purdue University}
  \streetaddress{1301 Third Street}
  \city{West Lafayette}
  \state{Indiana}
  \postcode{47906-4206}
}
\email{gould29@purdue.edu}



\begin{abstract}
Data Science has been a growing field since the 2010s. Between self-driving cars, movie and music recommendations, routing algorithms, and much more, both private companies and universities are making major investments in the field. As the field continues to mature, and we collect more data, the demand to store and analyze them will continue to increase. This increase in data availability and demand for analytics will put a strain on data centers and compute clusters—with implications for both energy costs and emissions. As the world battles a climate crisis, it is prudent for organizations with data centers to have a framework for combatting increasing energy costs and emissions to meet demand for analytics work.

In this paper, I present a generalized framework for organizations to audit data centers energy efficiency to understand the resources required to operate a given data center and effective steps organizations can take to improve data center efficiency and lower the environmental impact.
\end{abstract}




\maketitle





\begin{table*}[t]
    \centering
        \caption{Electrical consumption to train state-of-the-art natural language processing models, shown as powering average U.S. homes for 1 year}
        \label{tab:power_usage}
        \begin{tabular}{|c c c c|}
          \hline
          Model Name & GFLOP & Consumption & Number of Homes\\
          \hline
          \verb|T5-Small| & $1.8e^{11}$ & 10,666,034.61 kWh & 987.60\\
          \verb|T5-Base| & $6.6e^{11}$ & 39,108,793.55 kWh & 3,621.18\\
          \verb|T5-Large| & $2.31e^{12}$ & 136,880,777.40 kWh & 12,674.15\\
          \verb|T5-3B| & $9e^{12}$ & 533,301,730.30 kWh & 49,379.76\\
          \verb|T5-11B| & $3.3e^{13}$ & 1,955,439,678 kWh & 181,059.23\\
          \verb|BERT-Base| & $1.64e^{11}$ & 9,717,942.64 kWh & 899.81\\
          \verb|BERT-Large| & $5.33E^{11}$ & 31,583,313.58 kWh & 2,924.38\\
          \verb|RoBERTa-Base| & $1.5e^{12}$& 88,883,621.71 kWh & 8,229.964\\
          \verb|RoBERTa-Large| & $4.26e^{12}$ & 252,429,485.70 kWh & 23,373.10\\
          \verb|GPT-3 Small| & $2.25e^{11}$ & 13,332,543.26 kWh & 1,234.49\\
          \verb|GPT-3 Medium| & $6.41e^{11}$ & 37,982,934.34 kWh & 3,516.94\\
          \verb|GPT-3 Large| & $1.37e^{12}$ & 81,180,374.50 kWh & 7,516.70\\
          \verb|GPT-3 XL| & $2.38e^{12}$ & 141,028,679.80 kWh & 13,058.21\\
          \verb|GPT-3 2.7B| & $4.77e^{12}$ & 282,649,917 kWh & 26,171.29\\
          \verb|GPT-3 6.7B| & $1.2e^{13} $& 711,068,973.70 kWh & 65,839.72\\
          \verb|GPT-3 13B| & $2.31e^{13}$ & 1,368,807,774 kWh & 126,741.46\\
          \verb|GPT-3 175B| & $3.14e^{14}$ & 18,606,304,812 kWh & 1,722,806\\
          \hline
        \end{tabular}
\end{table*}

\section{Introduction}

With the increased demand for access to computational resources, the amount of energy required to operate data centers is expected to continue to grow. In 2020, global data center usage was $> 200 TWh$~\cite{capgemini}. According to the Central Intelligence Agency’s 2020 Fact Book, in 2016, the electrical consumption of Poland, Sweden, and Norway were $142 TWh$, $127 TWh$, and $126 TWh$, respectively~\cite{consumption}. By 2030, global data center electricity usage is predicted to exceed 2 trillion $TWh$~\cite{capgemini}. In the midst of a global climate crisis, data centers’ emissions and contributions are non-trivial and should be taken seriously to mitigate. The need for a generalized framework for organizations to audit data centers energy efficiency is great.

Data center emissions has implications in the fields of Artificial Intelligence and Machine Learning, too—as training models, and running inference in a production environment, requires access to compute resources, such as graphic processing units (GPU). Open AI’s Tom B. Brown et al.’s work on GPT-3 made reference to energy usage and implications of training models; however, the paper did not offer solutions to this growing problem ~\cite{brown2020language}. The authors reported estimates of the compute resources required to train 17 recent state-of-the-art NLP models. The numbers provided are not accessible to non-subject matter experts and difficult to visualize. I took these estimates and converted the power required to the number of average U.S. homes that can be powered for one year, depicted in \autoref{tab:power_usage}. For the purposes of this analysis, I assume $16.876 GFLOP/watt$, per a conservative cursory search of the Top500 List ~\cite{November2020TOP500} and a similar analysis only for GPT-3 performed by Matthew Burruss~\cite{burruss}, and average U.S. home electrical consumption as $900 kWh$ per month, per the U.S. Energy Information Administration~\cite{us_usage}.

Given the recent popularity of Data Science, and the increased demand for computational resources, previous work has been completed in attempting to create a set of guidelines and standards for organizations to follow to ensure data center efficiency. For example, the Lawrence Berkeley National Laboratory’s Center of Expertise for Energy Efficiency in Data Centers updated its \textit{Data Center Master List of Energy Efficiency Measures} in September 2020~\cite{masterlist}. This document is a collection of best practices for energy efficient data center operations, split into 7 sections: \textit{Data Center Energy Efficiency Management}, \textit{IT Power Distribution Chain}, \textit{IT Equipment}, \textit{Lighting}, \textit{Air Management,} \textit{Cooling the Data Center Space}, and \textit{Central Cooling Plant}. While the report has a large quantity of best practices and action items, many of the recommendations provided are resource-intensive—which may put smaller organizations at a disadvantage—and may not provide enough detail to be clearly understood and actionable. The Master List’s \textit{EEM 1-2.7: Perform an Energy Audit}, for example, does not provide enough detail to carry out an audit. The Master List does, however, link the \textit{DCEE Toolkit’s Energy Assessment Process Manual}~\cite{auditdoe}. The process manual falls victim to the similar shortfall of the Master List, whereby, quantity of content overshadows specificity and accessibility.

Another example of recent work is the U.S. Department of Energy’s \textit{Best Practices Guide for Energy-Efficient Data Center Design}~\cite{BestPracticesGuideforEnergyEfficien}. This guide is similar to the Master List, as it walks through best practices for various aspects of the data center, such as air management, cooling systems, environmental conditions, and IT systems. While the design guide offers more specificity than the Master List, it is not only nearly a decade old, but lacks the diversity and quantity of best practices provided by the Master List. Most apparent is the absence of guidelines to execute an energy efficiency audit. Furthermore, neither guide provides insight into organizational challenges associated with embarking on an audit, such as: creating an audit team, setting goals and benchmarks, and communication/goal accountability. I aim to address these important aspects of the audit process, and close these gaps, in section 3.

My primary goal for this paper is to leverage, enhance, and complement previous work in this space to provide a specific framework for an energy audit of a data center—with clear suggestions for organizations of varying size and resource availability. This is accomplished via the “full” and “lite” audit frameworks identified in this work.

The “full” audit is geared toward large, mature organizations with the resources required to not only perform deep analyses of many inner-workings of a data center, but the capital needed to address shortcomings identified in the audit.

For smaller organizations with fewer resources or expertise in high-performance computing, I offer a “lite” audit to focus effort on more attainable, less resource-intensive aspects of data center energy efficiency—as well as provide the roadmap to more mature energy assessment guidance to anticipate for future growth of the organization.


\section{Energy Audit}

This section contains the guidelines which comprise both the “full” and “lite” data center energy audits. The guidelines are binned into six categories, following the naming conventions established in the Master List~\cite{masterlist}: \textit{Cooling Air and Air Management}, \textit{Environmental Conditions}, \textit{Global}, \textit{IT Equipment}, \textit{IT Power Distribution Chain}, and \textit{Lighting}.

\subsection{"Full" vs. "Lite" Audit Frameworks}

The appendix offers two tables, depicting the components which make up the “full” and “lite” audits. To reiterate: the "full" audit is intended for large organizations with the resources required to undertake a laborious, and potentially expensive, audit. The "lite" audit framework is meant for smaller organizations with fewer resources, to serve as an accessible option to have the ability to gain an understanding of the state of data center energy efficiency. The action items of the "lite" audit framework generally require fewer resources and expertise than large-scale equipment replacement specified in the "full" audit. To conduct a "lite" audit, only complete items listed in \autoref{tab:lite}. To conduct a "full" audit, an organization must complete items all "lite" (\autoref{tab:lite}) and "full" (\autoref{tab:full}) options, depicted in \autoref{tab:all}.

\subsection{Audit Item Structure}

All entries in the proceeding subsections will use the following structure:
\begin{itemize}
\item \textbf{Audit Level}: At what level (e.g., individual rack, equipment room, entire data center, etc.) shall the analysis cover
\item \textbf{Item Description}: Description of what the audit item is and how to perform it.
\item \textbf{Metric}: How to measure the audit item.
\item \textbf{Desired Output(s) \& Goal(s)}: 
\begin{enumerate}
    \item Takeaway(s) from the audit item.
\end{enumerate}
\item \textbf{Action(s) to Take}: 
\begin{enumerate}
    \item What action(s) should follow completion of this audit item (i.e., mitigation strategies to improve efficiency).
\end{enumerate}
\item \textbf{Audit Type}: "Full" or "Lite" item.
\end{itemize}

\subsection{Cooling Air and Air Management}

Items related to efficiently managing airflow and cool air within the data center.

\subsubsection{Return Temperature Index ($RTI$)\\}
The Return Temperature Index ($RTI$)~\cite{rti_metric} is a key indicator explaining how effectively of the equipment room air-management system is operating. The $RTI$ reflects either net by-pass air ($RTI < 100\%$) or net recirculation air ($RTI > 100\%$).
\begin{itemize}
    \item \textbf{Audit Level}: Perform at the equipment room or entire data center level.
    \item \textbf{Item Description}: Capture the $RTI$ for either an equipment room in a multi-room data center, or roll up to the entire data center level.
    \item \textbf{Metric}: Percentage: (measure of net by-pass air ($RTI < 100\%$) or net recirculation air ($RTI > 100\%$))
    \item \textbf{Desired Output(s) \& Goal(s)}:
    \begin{enumerate}
        \item Identify which rooms or data centers do not have sufficient airflow management (i.e., $RTI \ne 100\%$).
    \end{enumerate}
    \item \textbf{Action(s) to Take}:
    \begin{enumerate}
        \item Adjust room air management to ensure optimal $RTI$ ($100\%$).
    \end{enumerate}
    \item \textbf{Audit Type}: Full.
\end{itemize}

\subsubsection{Alternating Hot and Cold Rack Aisles\\}
Aisles of server racks should alternate temperature between hot and cold. This can be achieved by ensuring the equipment is installed into the racks to create a front-to-back pattern of airflow, which will bring in cold air from the HVAC-chilled cold aisles and get rid of hot air via the hot aisles located behind the racks ~\cite{aisles}.
\begin{itemize}
    \item \textbf{Audit Level}: Perform at the individual aisle level.
    \item \textbf{Item Description}: Determine how many aisles of racks are alternating hot and cold temperatures.
    \item \textbf{Metric}: Percentage of the aisles compliant with the alternating hot and cold aisle scheme: $COMPLIANCE = \frac{\#\ compliant\ aisles}{\#\ total\ aisles}$. 
    \item \textbf{Desired Output(s) \& Goal(s)}:
    \begin{enumerate}
        \item Identify which rows do not follow the alternating hot and cold aisle scheme.
    \end{enumerate}
    \item \textbf{Action(s) to Take}:
    \begin{enumerate}
        \item Re-organize server racks and HVAC system to ensure $100\%$ compliance.
    \end{enumerate}
    \item \textbf{Audit Type}: Full.
\end{itemize}

\subsubsection{Physical Separation of Hot and Cold Air of Rack Aisles\\}
To further reduce mixing the cold supply air with the hot exhaust air (e.g., using flexible strip curtains or rigid enclosures). This, in turn, will reduce fan energy and improve chiller efficiency~\cite{masterlist}.According to the U.S. Department of Energy~\cite{ BestPracticesGuideforEnergyEfficien}, this can reduce fan energy requirements by up to $20\%-25\%$.
\begin{itemize}
    \item \textbf{Audit Level}: Perform at the individual aisle level.
    \item \textbf{Item Description}: Determine how many aisles of racks provide physical barriers to other aisles.
    \item \textbf{Metric}: Percentage of the aisles compliant physical barrier scheme: $COMPLIANCE = \frac{\#\ compliant\ aisles}{\#\ total\ aisles}$. 
    \item \textbf{Desired Output(s) \& Goal(s)}:
    \begin{enumerate}
        \item Identify which aisles do not follow the physical barrier scheme.
    \end{enumerate}
    \item \textbf{Action(s) to Take}:
    \begin{enumerate}
        \item With non-compliant aisles, install a barrier (e.g., strip curtain, rigid enclosures, etc.) to ensure $100\%$ compliance.
    \end{enumerate}
    \item \textbf{Audit Type}: Lite.
\end{itemize}

\begin{figure}
    \centering
    \includegraphics[width=\linewidth]{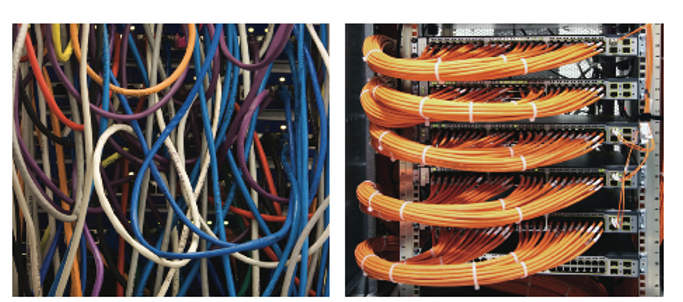}
    \caption{An example of unstructured cabling (left) and structured cabling (right)~\cite{cable_structure_image}.}
    \label{fig:cable_structure}
  \end{figure}

\subsubsection{Structured Cabling for a Rack\\}
Unstructured cable (see \autoref{fig:cable_structure}) can restrict airflow to and from the rack, allowing for neither cold air to cool the server nor hot air to expel. Use structured cabling (see\autoref{fig:cable_structure}) to avoid this.
\begin{itemize}
    \item \textbf{Audit Level}: Perform at the individual rack level.
    \item \textbf{Item Description}: Determine how many racks within the data center follow acceptable structured cabling best practices.
    \item \textbf{Metric}: Percentage of the racks leveraging structured cabling: $COMPLIANCE = \frac{\#\ compliant\ racks}{\#\ total\ racks}$. 
    \item \textbf{Desired Output(s) \& Goal(s)}:
    \begin{enumerate}
        \item Identify which racks have unstructured cabling.
    \end{enumerate}
    \item \textbf{Action(s) to Take}:
    \begin{enumerate}
        \item With non-compliant racks, re-wire to ensure structured cabling.
    \end{enumerate}
    \item \textbf{Audit Type}: Lite.
\end{itemize}

\subsubsection{Air Filter MERV Rating Compliance\\}
The Minimum Efficiency Reporting Values (MERV) rating is the measure of an air filter’s ability to capture larger particles between 0.3 and 10 microns ($\upmu$m)~\cite{WhatisaMERVratingIndoorAirQual}. According to the American Society of Heating (ASHRAE), Refrigerating and Air-Conditioning Engineers~\cite{gaseous}, recommends that any air entering the data center use MERV 11-13 filters, whereas the air continuously present in the data center can use MERV 8 filters.
\begin{itemize}
    \item \textbf{Audit Level}: Perform at the individual filter level.
    \item \textbf{Item Description}: Determine how many air filters within the data center are compliant $w.r.t.$ MERV rating, for an air filter's given purpose (i.e., external air vs. continuous air internal to the data center).
    \item \textbf{Metric}: Percentage of the compliant air filters: $COMPLIANCE = \frac{\#\ compliant\ air\ filters}{\#\ total\ air\ filters}$. 
    \item \textbf{Desired Output(s) \& Goal(s)}:
    \begin{enumerate}
        \item Identify which air filters are not compliant.
    \end{enumerate}
    \item \textbf{Action(s) to Take}:
    \begin{enumerate}
        \item With non-compliant air filters, replace with the appropriate MERV-rated air filter.
    \end{enumerate}
    \item \textbf{Audit Type}: Lite.
\end{itemize}

\begin{table*}[t]
    \centering
        \caption{Data center ambient temperature ranges suggested by ASHRAE~\cite{BestPracticesGuideforEnergyEfficien}}
        \label{tab:temp}
        \begin{tabular}{|c c c c|}
          \hline
          Temperature & Class 1 and Class 2 Recommended Range & Class 1 Allowable Range & Class 2 Allowable Range\\
          \hline
          \verb|Low Temperature Limit| & 64.4\degree F & 59\degree F & 50\degree F\\
          \verb|High Temperature Limit| & 80.6\degree F & 89.6\degree F & 95\degree F\\
          \hline
        \end{tabular}
\end{table*}

\subsection{Environmental Conditions}

Items related to environmental factors, such as: data center ambient temperature, rack cooling, etc.

\subsubsection{Data Center Ambient Temperature\\}
One of the simplest, yet effective, measures an organization can take to immediately improve a data center’s energy efficiency is analyze the data center’s ambient temperature. According to a study conducted by the U.S. Environmental Protection Agency and the U.S. Department of Energy~\cite{cable_structure_image}, many data centers set their temperatures much lower than the suggested range. Some data centers in the study were reported to have ambient temperatures as low as 55\degree F. The suggested range by ASHRAE~\cite{BestPracticesGuideforEnergyEfficien} is depicted in \autoref{tab:temp}. The agencies also report that data centers can save up to $4-5\%$ in energy costs for every 1\degree F increase in server inlet temperature.
\begin{itemize}
    \item \textbf{Audit Level}: Perform at the entire data center level.
    \item \textbf{Item Description}: Capture hourly data center ambient temperature ranges.
    \item \textbf{Metric}:
    \begin{enumerate}
        \item Percentage of the time within the suggested ASHRAE range: $COMPLIANCE = \frac{\#\ acceptable\ ambient\ temperatue\ readings}{\#\ total\ ambient\ temperature\ readings}$.
        \item Plot the ambient temperature over time, and aim to be near the high temperature range, depicted in \autoref{tab:temp}.
    \end{enumerate}
    \item \textbf{Desired Output(s) \& Goal(s)}:
    \begin{enumerate}
        \item Determine if data center follows ASHRAE guidelines for data center temperature ranges.
    \end{enumerate}
    \item \textbf{Action(s) to Take}:
    \begin{enumerate}
        \item Raise the data center's ambient temperature to not only be within the acceptable ASHRAE suggested temperature range, but strive to be near the high temperature limit.
    \end{enumerate}
    \item \textbf{Audit Type}: Lite.
\end{itemize}

\subsubsection{Rack Cooling Index($RCI$)\\}
The Rack Cooling Index ($RCI$)~\cite{rci_metric} is a key indicator explaining how effectively individual racks are cooled and maintained. The $RCI$ comprises of two metrics: $RCI_{HI}$ and $RCI_{LO}$, where $RCI_{HI}$ measures the presence or absence of over-temperatures and $RCI_{LO}$ measures the presence or absence of under-temperatures.
\begin{itemize}
    \item \textbf{Audit Level}: Perform at the individual rack level.
    \item \textbf{Item Description}: Capture the $RCI_{HI}$ and $RCI_{LO}$ for a given rack.
    \item \textbf{Metric}: Percentage for both $RCI_{HI}$ and $RCI_{LO}$.
    \item \textbf{Desired Output(s) \& Goal(s)}:
    \begin{enumerate}
        \item Identify which rooms or data centers do not have sufficient temperature management (i.e., $RCI_{type} \ne 100\%$).
    \end{enumerate}
    \item \textbf{Action(s) to Take}:
    \begin{enumerate}
        \item Adjust rack temperature management to ensure optimal $RCI_{HI}$ and $RCI_{LO}$ ($100\%$).
    \end{enumerate}
    \item \textbf{Audit Type}: Full.
\end{itemize}

\begin{table*}[t]
    \centering
        \caption{U.S. Department of Energy recommendations~\cite{BestPracticesGuideforEnergyEfficien} for standard data center energy efficiency metrics}
        \label{tab:doe}
        \begin{tabular}{|c c c c|}
          \hline
          Metric Name & Standard Score & Good Score & Better Score\\
          \hline
          \verb|Power Usage Effectiveness (PUE)| & 2 & 1.4 & 1.1\\
          \verb|Data Center Infrastructure Efficiency (DCIE)| & 0.5 & 0.7 & 0.9\\
          \verb|HVAC System Effectivenss (HVACSE)| & 0.7 & 1.4 & 2.5\\
          \verb|Airflow Efficiency (AE)| & 1.25 $W/cfm$ & 0.75 $W/cfm$ & 0.5 $kW/cfm$\\
          \verb|Cooling System Efficiency (CSE)| & 1.1 $kW/ton$ & 0.8 $kW/ton$ & 0.6 $kW/ton$\\
          \hline
        \end{tabular}
\end{table*}

\subsection{Global}

High-level items to monitor overall efficiency of energy use, such as; electrical, water, HVAC, airflow, etc.

\subsubsection{Power Usage Effectiveness ($PUE$)\\}
The Power Usage Effectiveness ($PUE$)~\cite{BestPracticesGuideforEnergyEfficien} is a standard metric to measure the overall efficiency of power used by the data center.
\begin{itemize}
    \item \textbf{Audit Level}: Perform at the entire data center level.
    \item \textbf{Item Description}: Capture the $PUE$ at an organization-defined cadence to track over time.
    \item \textbf{Metric}: The ratio of the total power to run the data center facility to the total power drawn by all IT equipment: $PUE = \frac{Total\ Facility\ Power}{IT\ Equipment\ Power}$.
    \item \textbf{Desired Output(s) \& Goal(s)}:
    \begin{enumerate}
        \item Benchmark $PUE$ per data center; overall current state of data center energy efficiency, per U.S. Department of Energy standards, depicted in \autoref{tab:doe}.
    \end{enumerate}
    \item \textbf{Action(s) to Take}:
    \begin{enumerate}
        \item Once an organization has a benchmark understanding of its $PUE$, it can use the non-metric best practices to take actions to improve $PUE$ and track it over time. This benchmark can also be used in coöperation with any existing sustainability strategy from the organization, to drive goals for future $PUE$ values (see section 3).
    \end{enumerate}
    \item \textbf{Audit Type}: Lite.
\end{itemize}

\subsubsection{Data Center Infrastructure Effectiveness ($DCIE$)\\}
The Data Center Infrastructure Effectiveness ($DCIE$)~\cite{BestPracticesGuideforEnergyEfficien} is the inverse of $PUE$-defined as the ratio of the total power drawn by all IT equipment to the total power to run the data center facility.
\begin{itemize}
    \item \textbf{Audit Level}: Perform at the entire data center level.
    \item \textbf{Item Description}: Capture the $DCIE$ at an organization-defined cadence to track over time.
    \item \textbf{Metric}: The ratio of the total power drawn by all IT equipment to the total power to run the data center facility (inverse of $PUE$): $DCIE = \frac{1}{PUE} = \frac{Total\ Facility\ Power}{IT\ Equipment\ Power}$.
    \item \textbf{Desired Output(s) \& Goal(s)}:
    \begin{enumerate}
        \item Benchmark $DCIE$ per data center; overall current state of data center energy efficiency, per U.S. Department of Energy standards, depicted in \autoref{tab:doe}.
    \end{enumerate}
    \item \textbf{Action(s) to Take}:
    \begin{enumerate}
        \item Once an organization has a benchmark understanding of its $DCIE$, it can use the non-metric best practices to take actions to improve $DCIE$ and track it over time. This benchmark can also be used in coöperation with any existing sustainability strategy from the organization, to drive goals for future $DCIE$ values (see section 3).
    \end{enumerate}
    \item \textbf{Audit Type}: Lite.
\end{itemize}

\subsubsection{Energy Reuse Effectiveness ($ERE$)\\}
The Energy Reuse Effectiveness ($ERE$)~\cite{BestPracticesGuideforEnergyEfficien} is the ratio of the total energy to run the data center facility minus the reuse energy to the total energy drawn by all IT equipment. $ERE$ value range: $0-\infty$. For example, an $ERE$ of $0$ means that $100\%$ of the energy brought into the data center is reused elsewhere, outside of the data center control volume.
\begin{itemize}
    \item \textbf{Audit Level}: Perform at the entire data center level.
    \item \textbf{Item Description}: Capture the $ERE$ at an organization-defined cadence to track over time.
    \item \textbf{Metric}: The ratio of the total energy to run the data center facility minus the reuse energy to the total energy drawn by all IT equipment: $ERE = \frac{Cooling\ +\ Power\ +\ Lighting\ +\ IT\ -\ Reuse\ Energy}{IT\ Equipment\ Energy}$.
    \item \textbf{Desired Output(s) \& Goal(s)}:
    \begin{enumerate}
        \item Benchmark $ERE$ per data center; overall current state of data center energy efficiency.
    \end{enumerate}
    \item \textbf{Action(s) to Take}:
    \begin{enumerate}
        \item Once an organization has a benchmark understanding of its $ERE$, it can use the non-metric best practices to take actions to improve $ERE$ and track it over time. This benchmark can also be used in coöperation with any existing sustainability strategy from the organization, to drive goals for future $ERE$ values (see section 3).
    \end{enumerate}
    \item \textbf{Audit Type}: Lite.
\end{itemize}

\subsubsection{HVAC System Effectiveness ($HVACSE$)\\}
HVAC System Effectiveness ($HVACSE$)~\cite{BestPracticesGuideforEnergyEfficien} is the ratio of the annual IT equipment energy to the annual HVAC system energy.
\begin{itemize}
    \item \textbf{Audit Level}: Perform at the entire data center level.
    \item \textbf{Item Description}: Capture the $HVACSE$ annually to track over time.
    \item \textbf{Metric}: The ratio of the annual IT equipment energy to the annual HVAC system energy: $HVACSE = \frac{kWh/yr_{IT}}{kWh/yr_{HVAC}}$.
    \item \textbf{Desired Output(s) \& Goal(s)}:
    \begin{enumerate}
        \item Benchmark $HVACSE$ per data center; overall current state of data center energy efficiency, per U.S. Department of Energy standards, depicted in \autoref{tab:doe}.
    \end{enumerate}
    \item \textbf{Action(s) to Take}:
    \begin{enumerate}
        \item Once an organization has a benchmark understanding of its $HVACSE$, it can use the non-metric best practices to take actions to improve $HVACSE$ and track it over time. This benchmark can also be used in coöperation with any existing sustainability strategy from the organization, to drive goals for future $HVACSE$ values (see section 3).
    \end{enumerate}
    \item \textbf{Audit Type}: Lite.
\end{itemize}

\subsubsection{Airflow Efficiency ($AE$)\\}
Airflow Efficiency ($AE$)~\cite{BestPracticesGuideforEnergyEfficien} is the overall measure of how efficiently air is moved through the data center.
\begin{itemize}
    \item \textbf{Audit Level}: Perform at the individual fan level.
    \item \textbf{Item Description}: Capture the $AE$ at an organization-defined cadence to track over time; roll up to the entire data center level.
    \item \textbf{Metric}: The overall airflow efficiency in terms of the total fan power required per unit of airflow: $AE = \frac{Total\ Fan\ Power\ (W)}{Total\ Fan\ Airflow\ (cfm)}$.
    \item \textbf{Desired Output(s) \& Goal(s)}:
    \begin{enumerate}
        \item Benchmark $AE$ per data center; overall current state of data center energy efficiency, per U.S. Department of Energy standards, depicted in \autoref{tab:doe}.
    \end{enumerate}
    \item \textbf{Action(s) to Take}:
    \begin{enumerate}
        \item Once an organization has a benchmark understanding of its $AE$, it can use the non-metric best practices to take actions to improve $AE$ and track it over time. This benchmark can also be used in coöperation with any existing sustainability strategy from the organization, to drive goals for future $AE$ values (see section 3).
    \end{enumerate}
    \item \textbf{Audit Type}: Lite.
\end{itemize}

\subsubsection{Cooling System Efficiency ($CSE$)\\}
Cooling System Efficiency ($CSE$)~\cite{BestPracticesGuideforEnergyEfficien} is a metric to measure the efficiency of an HVAC system.
\begin{itemize}
    \item \textbf{Audit Level}: Perform at the individual HVAC system level.
    \item \textbf{Item Description}: Capture the $CSE$ at an organization-defined cadence to track over time; roll up to the entire data center level.
    \item \textbf{Metric}: The ratio of average cooling system power usage ($kW$) to the average data center cooling load (tons): $CSE = \frac{\mu\ Cooling\ System\ Power\ (kW)}{\mu\ Cooling\ Load\ (cfm)}$.
    \item \textbf{Desired Output(s) \& Goal(s)}:
    \begin{enumerate}
        \item Benchmark $CSE$ per data center; overall current state of data center energy efficiency, per U.S. Department of Energy standards, depicted in \autoref{tab:doe}.
    \end{enumerate}
    \item \textbf{Action(s) to Take}:
    \begin{enumerate}
        \item Once an organization has a benchmark understanding of its $CSE$, it can use the non-metric best practices to take actions to improve $CSE$ and track it over time. This benchmark can also be used in coöperation with any existing sustainability strategy from the organization, to drive goals for future $CSE$ values (see section 3).
    \end{enumerate}
    \item \textbf{Audit Type}: Lite.
\end{itemize}

\subsection{IT Equipment}

Items related to the performance and utilization of servers and other IT equipment.

\subsubsection{Identify Unused Operational Servers\\}
Running unused operational servers-or any IT equipment-can be expensive and wasteful of resources.
\begin{itemize}
    \item \textbf{Audit Level}: Perform at the entire data center level.
    \item \textbf{Item Description}: Determine which equipment are not used within the data center.
    \item \textbf{Metric}: Percentage of the equipment unused: $\frac{\#\ unused\ equipment}{\#\ total\ equipment}$.
    \item \textbf{Desired Output(s) \& Goal(s)}:
    \begin{enumerate}
        \item Identify which operational servers are not in use and consider them candidates for retirement.
    \end{enumerate}
    \item \textbf{Action(s) to Take}:
    \begin{enumerate}
        \item Determine which unused operational servers to retire, to ensure all servers that are drawing energy are in use.
    \end{enumerate}
    \item \textbf{Audit Type}: Lite.
\end{itemize}

\subsubsection{Monitor Server CPU Utilization\\}
Understand server performance $w.r.t.$ CPU utilization. Ignacio Cano et al.~\cite{cano2018adares} performed a study in which  80\% of devices in a sample of VMs have a maximum resource usage that is $<60\%$, leaving $40\%$ of allocated CPU resources un-utilized. This is a waste of resources, and can be corrected by optimizing server allocation.
\begin{itemize}
    \item \textbf{Audit Level}: Perform at the individual server level.
    \item \textbf{Item Description}: Determine if a server is under- or over-utilized:
    \begin{itemize}
        \item \textbf{Under-utilization}~\cite{cano2018adares}: $usage <= 50\%$
        \item \textbf{Correct utilization}~\cite{cano2018adares}: $51\% <= usage <= 85\%$
        \item \textbf{Over-utilization}~\cite{cano2018adares}: $usage <= 86\%$
    \end{itemize}
    \item \textbf{Metric}: Percentage of the CPU utilized.
    \item \textbf{Desired Output(s) \& Goal(s)}:
    \begin{enumerate}
        \item Identify which operational servers are under-utilized or over-utilized.
    \end{enumerate}
    \item \textbf{Action(s) to Take}:
    \begin{enumerate}
        \item Optimize server performance to avoid under-utilization (poor energy management) and over-utilization (wear and tear).
        \item Consolidate under-utilized servers.
    \end{enumerate}
    \item \textbf{Audit Type}: Full.
\end{itemize}

\subsubsection{Equipment Efficiency\\}
Understand how performant current IT equipment is $w.r.t.$ standard metric~\cite{November2020TOP500} of $GFLOP/W$.
\begin{itemize}
    \item \textbf{Audit Level}: Perform at the individual server level.
    \item \textbf{Item Description}: Determine how energy efficient a server is.
    \item \textbf{Metric}: $GFLOP/W$.
    \item \textbf{Desired Output(s) \& Goal(s)}:
    \begin{enumerate}
        \item Identify which equipment have a low value of $GFLOP/W$.
    \end{enumerate}
    \item \textbf{Action(s) to Take}:
    \begin{enumerate}
        \item Procure energy efficient servers to replace inefficient equipment.
    \end{enumerate}
    \item \textbf{Audit Type}: Lite.
\end{itemize}

\subsection{IT Power Distribution Chain}

Items related to the efficiency of systems providing the data center with power.

\subsubsection{Trace Data Center Power Generation Sources\\}
Understand from where power fed to the data center are coming. Over time, non-renewable power generation sources should be phased out in favor of replacing with renewable sources.
\begin{itemize}
    \item \textbf{Audit Level}: Perform at the entire data center level.
    \item \textbf{Item Description}: Determine if an energy supply for the data center is renewable or non-renewable, and how much power is coming from each source.
    \item \textbf{Metric}: Percentage of power from renewable vs. non-renewable sources: $\frac{Total\ kWh\ from\ renewable}{Total\ kWh\ drawn}$
    \item \textbf{Desired Output(s) \& Goal(s)}:
    \begin{enumerate}
        \item Identify how electrical power is generated for the data center.
    \end{enumerate}
    \item \textbf{Action(s) to Take}:
    \begin{enumerate}
        \item Work with the organization's facilities group to create a strategy to phase out non-renewable sources of power in favor of renewable sources. The goal is $100\%$.
    \end{enumerate}
    \item \textbf{Audit Type}: Lite.
\end{itemize}

\subsection{Lighting}

Items related understanding the impact of lights, lamps, and bulbs throughout the data center.

\subsubsection{LED Bulbs\\}
A quick and easy way an organization can reduce a data center's environmental footprint is by looking at the bulbs used in its lights and lamps. According to the U.S. Department of Energy~\cite{LEDLightingDepartmentofEnergy}, LED bulbs required $75\%$ less energy than traditional incandescents.
\begin{itemize}
    \item \textbf{Audit Level}: Perform at the entire data center level.
    \item \textbf{Item Description}: Determine how many light bulbs in the data center are LEDs.
    \item \textbf{Metric}: Percentage of LED bulbs vs. traditional incandescents: $\frac{\#\ LED\ bulbs}{\#\ Total\ bulbs}$
    \item \textbf{Desired Output(s) \& Goal(s)}:
    \begin{enumerate}
        \item Identify which lamps do not contain LED bulbs.
    \end{enumerate}
    \item \textbf{Action(s) to Take}:
    \begin{enumerate}
        \item Replace all non-LED bulbs with LED. The goal is $100\%$.
    \end{enumerate}
    \item \textbf{Audit Type}: Lite.
\end{itemize}

\subsubsection{Lighting Control and Dimming\\}
Outside of replacing all non-LED bulbs with LEDs, one way an organization can reduce energy costs within a data center is to enable lights and lamps to operate at a lower level.
\begin{itemize}
    \item \textbf{Audit Level}: Perform at the entire data center level.
    \item \textbf{Item Description}: Determine how many lamps in the data center offer dimming.
    \item \textbf{Metric}: Percentage of lamps which support dimming: 
    \begin{itemize}
        \item $\frac{\#\ Lamps\ offer\ dimming}{\#\ Total\ lamps}$
    \end{itemize}
    \item \textbf{Desired Output(s) \& Goal(s)}:
    \begin{enumerate}
        \item Identify which lamps do not offer dimming.
    \end{enumerate}
    \item \textbf{Action(s) to Take}:
    \begin{enumerate}
        \item Install dimming controls on all non-dimming lamps and lights. The goal is $100\%$.
    \end{enumerate}
    \item \textbf{Audit Type}: Lite.
\end{itemize}

\subsubsection{Occupancy Sensors to Control Lights\\}
This is very closely related to the previous best practice. Sensors to determine if a person is present or absent, and turning lights off when not needed, can help save energy and costs in operating the data center.
\begin{itemize}
    \item \textbf{Audit Level}: Perform at the entire data center level.
    \item \textbf{Item Description}: Determine how many lamps in the data center offer control via occupancy sensors.
    \item \textbf{Metric}: Percentage of lamps which support control via occupancy sensors: $\frac{\#\ Lamps\ with\ sensors}{\#\ Total\ lamps}$
    \item \textbf{Desired Output(s) \& Goal(s)}:
    \begin{enumerate}
        \item Identify which lamps do not offer control via occupancy sensors.
    \end{enumerate}
    \item \textbf{Action(s) to Take}:
    \begin{enumerate}
        \item Install occupancy sensors on all non-equipped lamps and lights. The goal is $100\%$.
    \end{enumerate}
    \item \textbf{Audit Type}: Lite.
\end{itemize}

\section{Organizational Considerations}

Having a framework to conduct a data center energy efficiency audit is only the beginning in the journey for an organization to make its data centers more environmentally friendly. An organization must also have a team in place to execute the audit and make changes once benchmark performance data are collected and a strategy to make necessary changes to data centers-typically to contribute to an ongoing sustainability campaign (such as Google's~\cite{google}, Microsoft's~\cite{microsoft}, or Facebook's~\cite{facebook} commitments to become more sustainable organizations). The goal of this section is to provide the information necessary to help organizations in this space to ensure a successful energy audit, as well as enable the organization post-audit to make the changes necessary to improve energy efficiency and reduce the environmental impact.

\subsection{Creating a Team}
The Lawrence Berkeley National Laboratory’s Center of Expertise for Energy Efficiency in Data Centers provides some insight on the creation of a team to execute some of the best practices for energy-efficient data centers~\cite{masterlist}; however, given the ambiguity of the energy audit criterion described, I wanted to provide suggestions to complement the Lawrence Berkeley National Laboratory’s Center of Expertise for Energy Efficiency in Data Centers's suggestions~\cite{auditdoe}.

\begin{figure}
    \centering
    \includegraphics[width=\linewidth]{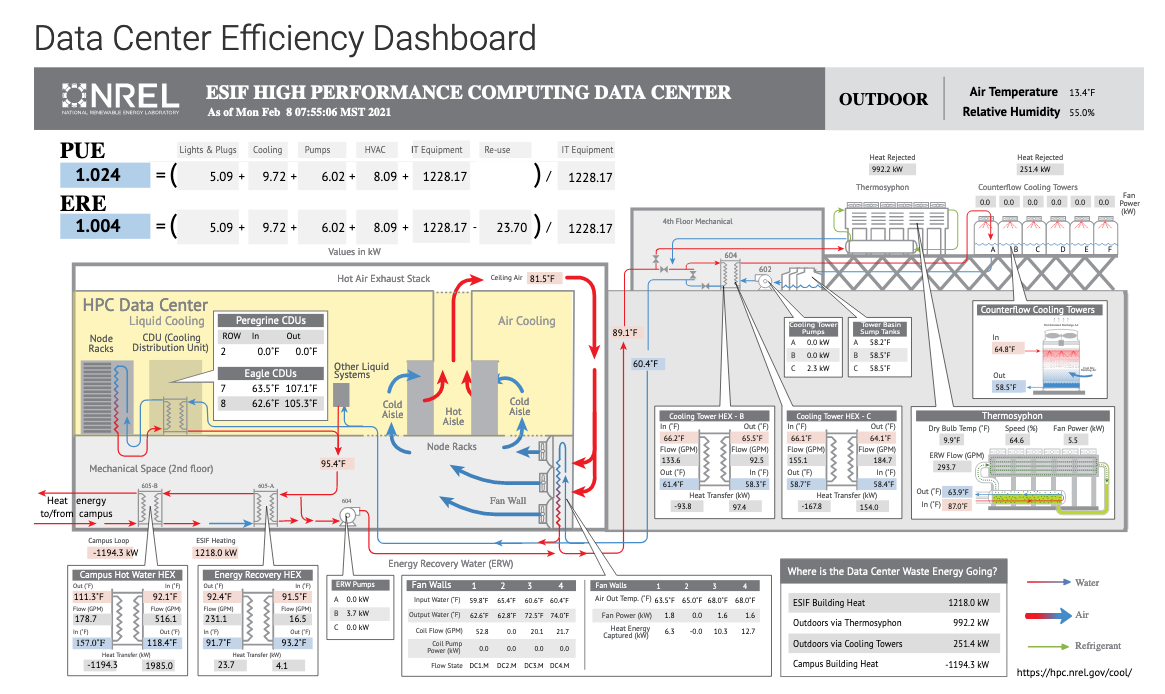}
    \caption{An example of a data center's energy efficiency dashboard, from the National Renewable Energy Laboratory's ESIF High Performance Computing Data Center~\cite{DataCenterEfficiencyDashboardHigh}.}
    \label{fig:dashboard}
  \end{figure}

Roles:
\begin{itemize}
    \item \textbf{Audit Lead}: The Audit Lead will coördinate with Site Lead(s) and Organization Engagement Lead/Liaison to organize audit and communicate results to responsible parties within the organization. The Audit Lead will also distribute work to the Audit Assessors and meet regularly to track audit progress and understand benchmark performance.
    \item \textbf{Site Lead(s)}: This is recommended by the Lawrence Berkeley National Laboratory’s Center of Expertise for Energy Efficiency in Data Centers~\cite{masterlist}. The lab defines this role as: "Data center technical representative that will be the primary person participating in the assessment." This individual will lead the Site Consulting Engineer(s) in any technical work that needs to be completed in the data center for the purposes of the audit, and will report findings back to the Audit Assessor(s) and Audit Lead.
    \item \textbf{Audit Assessor(s)}: This is recommended by the Lawrence Berkeley National Laboratory’s Center of Expertise for Energy Efficiency in Data Centers~\cite{masterlist}. The lab defines this role as: "The energy expert assigned to complete the...assessment; the expert serves as the facilitator for all activities." Depending on the size of the organization, $> 1$ Audit Assessor may be necessary-especially if multiple data centers are included in the audit.
    \item \textbf{Site Consulting Engineer(s)}: At least one Data Engineer per data center to provide information on infrastructure, installing sensors to collect data, handling operational servers, etc.
    \item \textbf{Data Scientist(s)}: At least one Data Scientist to collect, parse, visualize, and communicate data required to successfully execute the audit. Suggested action: create dashboards to report, track, and monitor data center performance over time-not only during the audit to benchmark performance. An example of this dashboard can be seen in \autoref{fig:dashboard}.
    \item \textbf{Organization Engagement Lead/Liaison}: Serves as the point of contact between organization management (e.g., for funding, marketing/communication/organizational publicity), facilities (e.g., for power generation, data center lighting control, etc.), and sustainability (e.g., for contributing to an organization's ongoing sustainability strategy).
\end{itemize}

\subsection{Setting Energy Efficiency Goals and Benchmarks Post-Audit}
Once the initial data center energy efficiency audit is complete, the organization will understand the environmental impact its data center(s) are having. At this time, it is prudent for the organization to make changes necessary to improve energy efficiency and reduce the environmental footprint of the data centers. Outside of the taking the recommended actions based on the result of the items in the audit (described in length in the previous section), the organization should continue to track and monitor its energy efficiency at a regular cadence (quarterly reporting is recommended, per~\cite{google}; however, a far more granular cadence of collecting these data can enable the organization to leverage techniques like machine learning to optimize the data center's parameters to maximize energy efficiency~\cite{optimize}).

An important consideration to make is setting annual goals for energy efficiency improvement, as well as reporting mechanisms to communicate these improvements. As seen in the examples from Google~\cite{google}, Microsoft~\cite{microsoft}, and Facebook~\cite{facebook}, these organizations have set ambitious environmental goals for the near future to reach carbon-free energy generation, carbon negativity, and net zero GHG emissions, respectively. However, smaller organizations-and higher education institutions-can set similar goals to improve its environmental footprint. For example, Purdue University is working toward its Sustainability Master Plan~\cite{purdue_sustainability}, consisting of goals related to energy, water, materials/waste, buildings, and the grounds of the University to reduce the impact operations have on the environment. To demonstrate that a data center energy efficiency initiative contributes to a strategy such as this:

Sustainability Master Plan Initiatives:
\begin{itemize}
    \item \textbf{Energy}~\cite{purdue_energy}:
    \begin{itemize}
        \item \textit{E-1 – Cut Carbon Emissions in Half}: "Reduce...carbon emissions by 50\% by FY25, with FY11 as the baseline year."
        \begin{enumerate}
            \item Improving data center energy efficiency (decreasing $PUE$, for example) will reduce the required power drawn from power sources potentially contributing to an organization's carbon emissions. Actions as trivial as dimming data center lights when not in use contributes to this goal.
            \item Over time, matriculating a data center's power generation sources from non-renewable sources to renewable sources can reduce an organization's overall carbon emission footprint.
        \end{enumerate}
        \item \textit{E-2 – No Net Gain}: "Cap total energy consumption at FY11 levels."
        \begin{enumerate}
            \item Improving data center energy efficiency (decreasing $PUE$, for example) will reduce the required power drawn from power sources potentially contributing to an organization's carbon emissions. Actions as trivial as dimming data center lights when not in use contributes to this goal.
            \item Over time, matriculating a data center's power generation sources from non-renewable sources to renewable sources can reduce an organization's overall carbon emission footprint.
        \end{enumerate}
    \end{itemize}
    \item \textbf{Water}~\cite{purdue_water}:
    \begin{itemize}
        \item \textit{W-1 - Reduce Water Consumption by 30\%}: "Reduce potable water consumption inside buildings and for irrigation by 30\% by FY25 on a gallon per square foot basis, with FY11 as the baseline year."
        \begin{enumerate}
            \item Employing best practices to improve rack cooling efficiency and optimizing operational servers in use can reduce the amount of chilling required for the data center, thus decreasing water consumption.
        \end{enumerate}
    \end{itemize}
\end{itemize}


\bibliographystyle{unsrt}
\bibliography{sample}

\clearpage


\begin{appendices}
\appendix

\begin{table*}[t]
    \centering
        \caption{Items included only in the "lite" data center energy efficiency audit.}
        \label{tab:lite}
        \begin{tabular}{|cc|}
          \hline
          Item Title & Audit Type\\
          \hline
          \verb|Physical Separation of Hot and Cold Air of Rack Aisles| & Lite\\
          \verb|Structured Cabling for a Rack| & Lite\\
          \verb|Air Filter MERV Rating Compliance| & Lite\\
          \verb|Data Center Ambient Temperature| & Lite\\
          \verb|Power Usage Effectiveness| & Lite\\
          \verb|Data Center Infrastructure Efficiency| & Lite\\
          \verb|Energy Reuse Effectiveness| & Lite\\
          \verb|HVAC System Effectiveness| & Lite\\
          \verb|Airflow Efficiency| & Lite\\
          \verb|Cooling System Efficiency| & Lite\\
          \verb|Identify Unused Operational Servers| & Lite\\
          \verb|Equipment Efficiency| &	Lite\\
          \verb|Identify Power Sources| & Lite\\
          \verb|LED Bulbs| & Lite\\
          \verb|Lighting Control and Dimming| & Lite\\
          \verb|Occupancy Sensors to Control Lights| & Lite\\
          \hline
        \end{tabular}
\end{table*}

\begin{table*}[t]
    \centering
        \caption{Items included only in the "full" data center energy efficiency audit. \textit{NOTE: To complete a "full" audit, the items from "lite" an "full" must both be completed.}}
        \label{tab:full}
        \begin{tabular}{|cc|}
          \hline
          Item Title & Audit Type\\
          \hline
          \verb|Return Temperature Index| & Full\\
          \verb|Alternating Hot and Cold Rack Aisles| & Full\\
          \verb|Rack Cooling Index| & Full\\
          \verb|Monitor Server CPU Utilization| & Full\\
          \hline
        \end{tabular}
\end{table*}

\begin{table*}[t]
    \centering
        \caption{All audit items, with their audit type classification. }
        \label{tab:all}
        \begin{tabular}{|cc|}
          \hline
          Item Title & Audit Type\\
          \hline
          \verb|Return Temperature Index| & Full\\
          \verb|Alternating Hot and Cold Rack Aisles| & Full\\
          \verb|Physical Separation of Hot and Cold Air of Rack Aisles| & Lite\\
          \verb|Structured Cabling for a Rack| & Lite\\
          \verb|Air Filter MERV Rating Compliance| & Lite\\
          \verb|Data Center Ambient Temperature| & Lite\\
          \verb|Rack Cooling Index| & Full\\
          \verb|Power Usage Effectiveness| & Lite\\
          \verb|Data Center Infrastructure Efficiency| & Lite\\
          \verb|Energy Reuse Effectiveness| & Lite\\
          \verb|HVAC System Effectiveness| & Lite\\
          \verb|Airflow Efficiency| & Lite\\
          \verb|Cooling System Efficiency| & Lite\\
          \verb|Identify Unused Operational Servers| & Lite\\
          \verb|Monitor Server CPU Utilization| & Full\\
          \verb|Equipment Efficiency| &	Lite\\
          \verb|Identify Power Sources| & Lite\\
          \verb|LED Bulbs| & Lite\\
          \verb|Lighting Control and Dimming| & Lite\\
          \verb|Occupancy Sensors to Control Lights| & Lite\\
          \hline
        \end{tabular}
\end{table*}

\end{appendices}

\end{document}